\documentclass[a4paper,fleqn,usenatbib,useAMS]{mnras}

\usepackage[T1]{fontenc}
\usepackage{ae,aecompl}

\usepackage{graphicx}	
\usepackage{amsmath}	
\usepackage{amssymb}	
\usepackage{epsfig,subfigure,amsmath}
\usepackage{array}
\usepackage{color}

\title[Jackknife resampling on mock catalogs]{Jackknife resampling technique on mocks: an alternative method for covariance matrix estimation}

\author[S.~ Escoffier et al.]{
S.~Escoffier$^{1}$\thanks{E-mail: escoffier@cppm.in2p3.fr},
M.-C.~Cousinou$^{1}$,
A.~Tilquin$^{1}$,
A.~Pisani$^{1,2,3}$,
A.~Aguichine$^{1,4}$,
\newauthor
S. de la Torre $^{5}$,
A. Ealet$^{1}$,
W. Gillard$^{1}$,
E. Jullo$^{5}$
\\
$^{1}$Aix Marseille Universit\'e, CNRS/IN2P3, CPPM UMR 7346, 13288, Marseille, France\\
$^{2}$Sorbonne Universit\'es, UPMC (Paris 06), UMR7095, Institut d'Astrophysique de Paris, 98bis Bd. Arago, F-75014, Paris, France\\
$^{3}$CNRS, UMR7095, Institut d'Astrophysique de Paris, 98bis Bd. Arago, F-75014, Paris, France\\
$^{4}$Paris-Saclay Universit\'e, ENS Cachan, 61 av. du Pr\'esident Wilson, 94230 Cachan, France\\
$^{5}$Aix Marseille Universit\'e, CNRS, LAM (Laboratoire d'Astrophysique de Marseille) UMR 7326, 13388, Marseille, France
}

\date{Accepted XXX. Received YYY; in original form ZZZ}

\pubyear{2015}

\begin{document}
\label{firstpage}
\pagerange{\pageref{firstpage}--\pageref{lastpage}}
\maketitle

\begin{abstract}
We present a fast and robust alternative method to compute covariance matrix in case of cosmology studies. Our method is based on the jackknife resampling applied on simulation mock catalogues. Using a set of 600 BOSS DR11 mock catalogues as a reference, we find that the jackknife technique gives a similar galaxy clustering covariance matrix estimate by requiring a smaller number of mocks. A comparison of convergence rates show that $\sim$7 times fewer simulations are needed to get a similar accuracy on variance. We expect this technique to be applied in any analysis where the number of available N-body simulations is low.
\end{abstract}




\section{Introduction}
During the last decade the observation of the large-scale structure of the universe has been intensified: the mapping of the three-dimensional distribution of galaxies reaches an unprecedented level of precision. However the accuracy on the parameters measurement is required at the sub-percent level to probe the nature of dark energy or modifications to gravity. Incoming and future large spectroscopic galaxy surveys such as the Dark Energy Spectroscopic Instrument~\citep{Levi:2013}, the Subaru Prime Focus Spectrograph~\citep{Ellis:2014} and the space-based Euclid mission~\citep{Laureijs:2011} aim to deliver high statistical accuracy. The challenge for these surveys is to control systematic biases and to contain all potential systematic errors within the bounds set by statistical ones.  

An essential step in galaxy clustering measurements is to get an unbiased estimate of the covariance matrix, which is needed to fully control the accuracy on the measured parameters. The difficulty to predict an accurate data covariance matrix relies intrinsically on the modeling of nonlinear gravitational evolution of matter distribution, galaxy bias and redshift-space distortions (RSD). The data covariance matrix can be estimated in a number of ways. A standard approach is to generate many mock samples that match the properties of the cosmological data set~\citep[e.g.][]{Manera:2013}. The data covariance matrix is computed from the scatter in the parameter value obtained for each statistically independent realization. Recent works~\citep{Taylor:2013, Dodelson:2013, Taylor:2014, Percival:2014} showed that this covariance estimate suffers from noise due to the finite number of mock realizations, implying an extra variance on estimated cosmological parameters, of $\mathcal{O}(1/(N_{mocks}-N_{bins}))$, where $N_{bins}$ is the total number of bins in the measurement. To keep this error contribution to an acceptable level one needs to estimate the covariance matrix from a very large number of mock realizations, satisfying $N_{mocks}\gg N_{bins}$. For instance in the SDSS survey, $600$ independent mock catalogs were generated for DR10 and DR11 BOSS measurements~\citep{Anderson:2014,Percival:2014}, and several tens of thousands are expected for the next generation of galaxy surveys. The main limitation of this approach is that the large number of mock realizations requires large computational resources, and this situation will be manageable with difficulty for surveys such as DESI or Euclid. 

Numerous efforts have been made to propose viable alternatives to this issue. Approximate methods aim at reducing computational time with fast gravity solvers, instead of producing full N-body simulations. Besides log-normal density field realizations~\citep{Coles:1991}, methods based on Lagrangian perturbation theory (LPT) have been proposed, such as \textsc{pthalos}~\citep{Scoccimarro:2002, Manera:2013}, \textsc{pinocchio}~\citep{Monaco:2002,Monaco:2013} and \textsc{patchy} algorithms~\citep{Kitaura:2015}. Other recent approaches include the effective Zel'dovich approximation \textsc{ezmocks}~\citep{Chuang:2015}, the quick particle mesh~\citep{White:2014} and the hybrid fast \textsc{cola} method~\citep{Tassev:2013,Koda:2015}.

Interestingly, statistical methods can be also applied in the limit of a small number of simulations, such as the shrinkage estimation~\citep{Pope:2008}, the resampling in Fourier modes~\citep{Schneider:2011} or the covariance tapering method~\citep{Paz:2015}, or using theoretically few-parameter models~\citep{Pearson:2016, Connell:2015}. 

Lastly, the data covariance matrix can be estimated through the resampling of data itself, using jackknife~\citep{Tukey:1958} or bootstrap~\citep{Efron:1979} methods. Such internal methods have been applied extensively in the past~\citep{Barrow:1984, Ling:1986, Hamilton:1993, Fisher:1994, Zehavi:2002}.
The jackknife resampling technique has some disadvantages~\citep{Norberg:2009}, but is usually discussed in the delete-one scheme based on the deleting single case from the original sample.

In this paper we propose an hydrid approach that deals with the jackknife resampling of mock catalogs (\textsc{smc} method) instead of the data itself. The sample covariance is estimated using the delete-$d$ jackknife scheme~\citep{Shao:1989} applied on an ensemble of independent realizations. The \textsc{smc} method can be particularly useful in case the number of available N-body light cone simulations is low. 

This paper is organized as follows. In Section~\ref{sec:cov} we present the formalism of covariance and precision matrix. We briefly remind how is estimated the sample covariance matrix from a set of independent simulations in Section~\ref{sec:cov_mock}. We next describe the jackknife resampling technique and give an estimate of the sample covariance matrix with our \textsc{smc} method in Section~\ref{sec:cov_jack}. In Section~\ref{sec:smc} we apply our method for the correlation function covariance matrix estimate. We describe in more detail our method of resampling mock realizations with the jackknife method and illustrate it with the sample mean and sample variance of the correlation function in Section~\ref{sec:smc_mean}. We discuss about the settings of jackknife parameters in the covariance matrix estimate in Section~\ref{sec:smc_param}. In Section~\ref{sec:smc_matrix} we present the comparison of the full covariance matrix obtained by the \textsc{smc} estimate to the method using simply N-body simulations. We next perform comparison of precision matrices in Section~\ref{sec:smc_precision}. Finally we compare the rates of convergence in Section~\ref{sec:convergence}. We conclude with a general discussion in Section~\ref{sec:conclusion}.

\section{Covariance and precision matrix}
\label{sec:cov}
We suppose here that the parameter estimation is derived from a likelihood analysis. If the data distribution is a multivariate Gaussian, then parameter constraints are obtained by minimizing:
\begin{equation}
   \chi^2=-2\ln{L}=\sum_{i,j=1}^{N_b} (x_{i}^{d}-x_{i}^{model}) \Psi^t_{ij} (x_{j}^{d}-x_{j}^{model})
	\label{eq:likelihood}
\end{equation}
where $x^d$ is the data collected in $N_b$ bins, $x^{model}$ is the model prediction and $\Psi^t$ is the inverse covariance matrix, the so-called precision matrix. The superscript $t$ denotes the true matrix.

\subsection{Independent mock realizations}
\label{sec:cov_mock}
\subsubsection{Covariance matrix}
The evaluation of the likelihood function requires the knowledge of the precision matrix $\Psi^t$, computed as the inverse of the covariance matrix. If a physical model is given for the covariance matrix, then the computation is analytical. Otherwise, as it is the case when statistical properties of the data are not well known, the data covariance matrix can be estimated from an ensemble of simulations. In such a case, an unbiased estimate of $C_{ij}$ is given by:
\begin{equation}
   \widehat{C}_{ij} = \frac {1} {N_s-1} \sum_{k=1}^{N_s} (y^k_i-\overline{y}_{i}) (y^k_j-\overline{y}_{j})
   \label{eq:cov_matrix}
\end{equation}
where $N_s$ is the number of independent realizations, $y_{i}^k$ is the value of the data parameter in bin $i$ for the $k$-th mock realization and $\overline{y}_{i}$ is the mean value in bin $i$ over the set of mock catalogs:
\begin{equation}
   \overline{y}_{i} = \frac {1} {N_s} \sum_{k=1}^{N_s} y^k_{i}
\end{equation}

The statistical properties of the sample covariance matrix $\widehat{C}$ of Eq.~(\ref{eq:cov_matrix}) follow a Wishart distribution~\citep{Wishart:1928}, which generalizes the $\chi^2$ distribution in case of multivariate Gaussian parameters. From the moments of the Wishart distribution one can deduce the variance of the elements of the sample covariance matrix:
\begin{equation}
   \sigma^2[ \widehat{C}_{ij}] = \frac {1} {N_s-1} [C_{ij}^2+C_{ii}C_{jj}]
    \label{eq:cov_matrix_var}
\end{equation}

\subsubsection{Precision matrix}
The precision matrix is defined as the inverse of the true covariance matrix ${\Psi}^t\equiv[C^t]^{-1}$. However, the measured $[\hat{C}]^{-1}$ is governed by the skewed inverse-Wishart distribution~\citep{Press:1982}, giving a biased estimate of the true precision matrix ${\Psi}^t$. 
If the condition $N_s>N_b+2$ is satisfied, where $N_b$ is the size of the data $x^d$, then an unbiased estimate of the precision matrix is given by~\citep{Hartlap:2007}:

\begin{equation}
   \widehat{\Psi} =  \frac {N_s-N_b-2} {N_s-1}  [\hat{C}]^{-1}
	\label{eq:hartlap_mean}
\end{equation}

\citet{Taylor:2013} and \citet{Dodelson:2013} found that the bias in the mean of the precision matrix is not the only effect that affects the accuracy of cosmological parameter estimation and that the covariance of the sample precision matrix should also be corrected, due to the finite number of realizations used. The unbiased variance on the elements of the precision matrix can be expressed as~\citep{Taylor:2013}:
\begin{equation}
   \sigma^2 [\widehat{\Psi}_{ij}] =  A \; [(N_s-N_b)\Psi_{ij}^2 + (N_s-N_b-2) \Psi_{ii} \Psi_{jj} ]
	\label{eq:hartlap_var}
\end{equation}
with
\begin{align}
   A = \frac {1} {(N_s-N_b-1)(N_s-N_b-4)} 
\end{align}

\subsection{The jackknife resampling on mock catalogs}
\label{sec:cov_jack}
\subsubsection{The delete-$d$ jackknife method}
The jackknife method is a statistical technique which aims at using observed data themselves to make an estimate of the error on the measurement \citep{Quenouille:1949, Tukey:1958}. 
Given a sample of size $N$, one can omit, in turn, each of the $N$ subsamples from the initial dataset to draw $N$ new datasets, each dataset being a $N-1$ sized sample (delete-$1$ method). The resulting mean and variance is computed over the entire ensemble of $N$ datasets. 

First attempts were done for 2-point clustering statistics from galaxy redshift surveys, where resampling technique was applied by removing each galaxy in turn from the catalog~\citep{Barrow:1984, Ling:1986}. However \citet{Norberg:2009} argued that, in order to avoid underestimating clustering uncertainties and to resolve computational troubles given the very high statistics in modern galaxy catalogs, the resampling of the data should be applied on $N_{s}$ sub-volumes into which the dataset has been split instead of on individual galaxies. The main issue is to be able to reproduce large-scale variance.
An other concern is about the delete-$1$ jackknife variance estimator, which is known to be asymptotically consistent for linear estimators (e.g. sample mean), but biased in non-linear cases (e.g. correlation function)~\citep{Miller:1974,Wu:1986}. 

Hence we consider in this paper the delete-{\it d} jackknife method proposed by \citet{Shao:1989}. The delete-{\it d} jackknife method consists of leaving out $N_d$ observations at a time instead of leaving out only one observation, which in our case means leaving out, in turn, $N_d$ subsamples amongst $N_s$ initial subsamples.  
The dimension of each new dataset is $(N_s-N_d)$ and the number of datasets is the combination $N_{\textsc{jk}}= { N_s \choose N_d}$. 
It can be shown that the delete-{\it d} jackknife method is asymptotically unbiased if $N_d$ satisfies both conditions $\frac{\sqrt{N_s}}{N_d} \rightarrow 0$ and $ N_s-N_d \rightarrow \infty $ \citep{Wu:1986}. This means it is preferable to choose 
\begin{equation}
\sqrt{N_s}<N_d<N_s
\label{eq:Nd}
\end{equation}

In the delete-{\it d} jackknife scheme where $N_{\textsc{jk}}$ jackknife resamplings are applied on a single realization, the covariance matrix is estimated by:
\begin{equation}
    \widehat{C}_{ij}=\frac{(N_s-N_d)}{N_d.N_{\textsc{jk}}} \sum\limits_{k=1}^{N_{\textsc{jk}}} (y_{i}^k-\overline{y}_{i}) (y_{j}^k-\overline{y}_{j}) 
 \label{eq:cov_matrix_JK}
\end{equation}
where $y_{i}^k$ is the value of the data parameter in the bin $i$ for the $k$-th jackknife configuration and $\overline{y}_{i}$ is the empirical average of the jackknife replicates:
\begin{equation}
  \overline{y}_{i}=\frac{1}{N_{\textsc{jk}}} \sum\limits_{k=1}^{N_{\textsc{jk}}} y_{i}^k
\end{equation}

\subsubsection{Average covariance and precision matrices}
The combined method we propose in this paper consists in averaging covariance matrices computed with $N_{\textsc{jk}}$ jackknife pseudo-realizations over a few sample of independent mock catalogs, and not just from a single realization. 

if ${}^{(m)}C_{ij}$ denotes the covariance matrix computed with $N_{\textsc{jk}}$ jackknife pseudo-realizations applied on the initial mock $m$ and satisfying Eq.~\ref{eq:cov_matrix_JK}, then we can define an estimator of the sample covariance matrix as the average of covariance matrices over $N_M$  independent mocks such as:
\begin{equation}
  \overline{C}_{ij} = \frac{1}{N_{M}}  \sum\limits_{m=1}^{N_{M}} {}^{(m)}\widehat{C}_{ij} 
 \label{eq:cov_matrix_average_JK}
\end{equation}

As $\overline{C}_{ij}$ is a sample mean value computed from uncorrelated ${}^{(m)}\widehat{C}_{ij}$, because independent mock catalogs are uncorrelated by definition, we are able to compute the variance on the sample mean as:
\begin{equation}
   \sigma^2[\widehat{C}_{ij}] = \frac{1}{N_{M}-1} \sum\limits_{m=1}^{N_M} ({}^{(m)}\widehat{C}_{ij}  - \overline{C}_{ij})^{2} 
   \label{eq:variance_average_JK}
\end{equation}

We define in this case the precision matrix as the inverse of the average covariance matrix:
\begin{equation}
 \widehat{\Psi}_{ij}=  [\overline{C}_{ij}]^{-1} 
 \label{eq:prec_matrix_JK}
 \end{equation}

The relevance of using an average covariance matrix over several mock realizations is discussed and highlighted in Sections~\ref{sec:smc_mean} and \ref{sec:smc_matrix}.

\section{Illustration using correlation function}
\label{sec:smc}
In this section we present the new hybrid method for the estimate of the covariance matrix applied on the two-point correlation function. We perform some validation tests and we use for comparison the estimate of the covariance matrix computed from $600$ independent mock realizations~\citep{Manera:2013} that we consider as a reference here.

\subsection{Method}
\label{sec:smc_method}
\subsubsection{Simulation data}
Galaxy mock catalogs used for this work have been generated for the SDSS-III/BOSS survey \citep{Dawson:2012} and used for the DR11 BOSS analysis~\citep{Anderson:2014}.They were generated using a method similar to \textsc{pthalos}~\citep{Scoccimarro:2002}, and a detailed description is given in \citet{Manera:2013}. We only consider the CMASS galaxy sample ($0.43<z<0.75$) for this study, with $600$ independent available realizations.

The study presented here has been restricted to the CFHT Legacy Survey W1 field in a multiparameters fit. We delay to a future paper the study extended to the full NGC and SGC BOSS footprints. 

In order to generate jackknife configurations, we divide our sample into separate regions on the sky of approximately equal area.   
Unless otherwise specified, we use for the following $N_s=12$ and $N_d=6$, cutting in the CFHTLS W1 field.

%



\subsubsection{Clustering statistics}
In the rest of this paper we consider two-point clustering statistics, by calculating the two dimensional correlation function $\xi(s,\mu)$ in which $s$ is the redshift-space separation of pairs of galaxies and randoms, and $\mu$ the cosine of the angle of the pair to the line-of-sight, using the \citet{Landy:1993} estimator:
\begin{equation}
\xi(s,\mu) = \frac {DD(s,\mu)-2DR(s,\mu)+RR(s,\mu)} {RR(s,\mu)}
\end{equation}
where $DD$, $DR$ and $RR$ represent the number of pairs of galaxies extracted from the galaxy sample $D$ and from the random sample $R$.

One galaxy sample $D$ is generated for each of the $N_{\textsc{jk}}$ jackknife combinations (924 combinations for the ${12 \choose 6}$ configuration), to which is associated one random catalog $R$ generated on the same deleted field with a number of randomly distributed points fifteen times larger than the number of data points.

We project the $\mu$-dependence to obtain multipoles of the correlation function:
\begin{equation}
\xi_l(s) = \frac{2l+1}{2} \int_{-1}^{1}d\mu \xi(s,\mu) L_l(\mu)
\end{equation}
where $L_l(\mu)$ is the Legendre polynomial of order $l$.
Monopole $\xi_0(s)$ and quadrupole $\xi_2(s)$ components are then projected on $2 \times 11$ bins of equal $\Delta{\log{(s)}} = 0.1$ $h^{-1}$Mpc width, varying from $s = 2.8$ $h^{-1}$Mpc to $s=28.2$ $h^{-1}$Mpc.

\subsection{Sample mean and variance}
\label{sec:smc_mean}
\begin{figure}
   \begin{center}
      \includegraphics[height=2.5in]{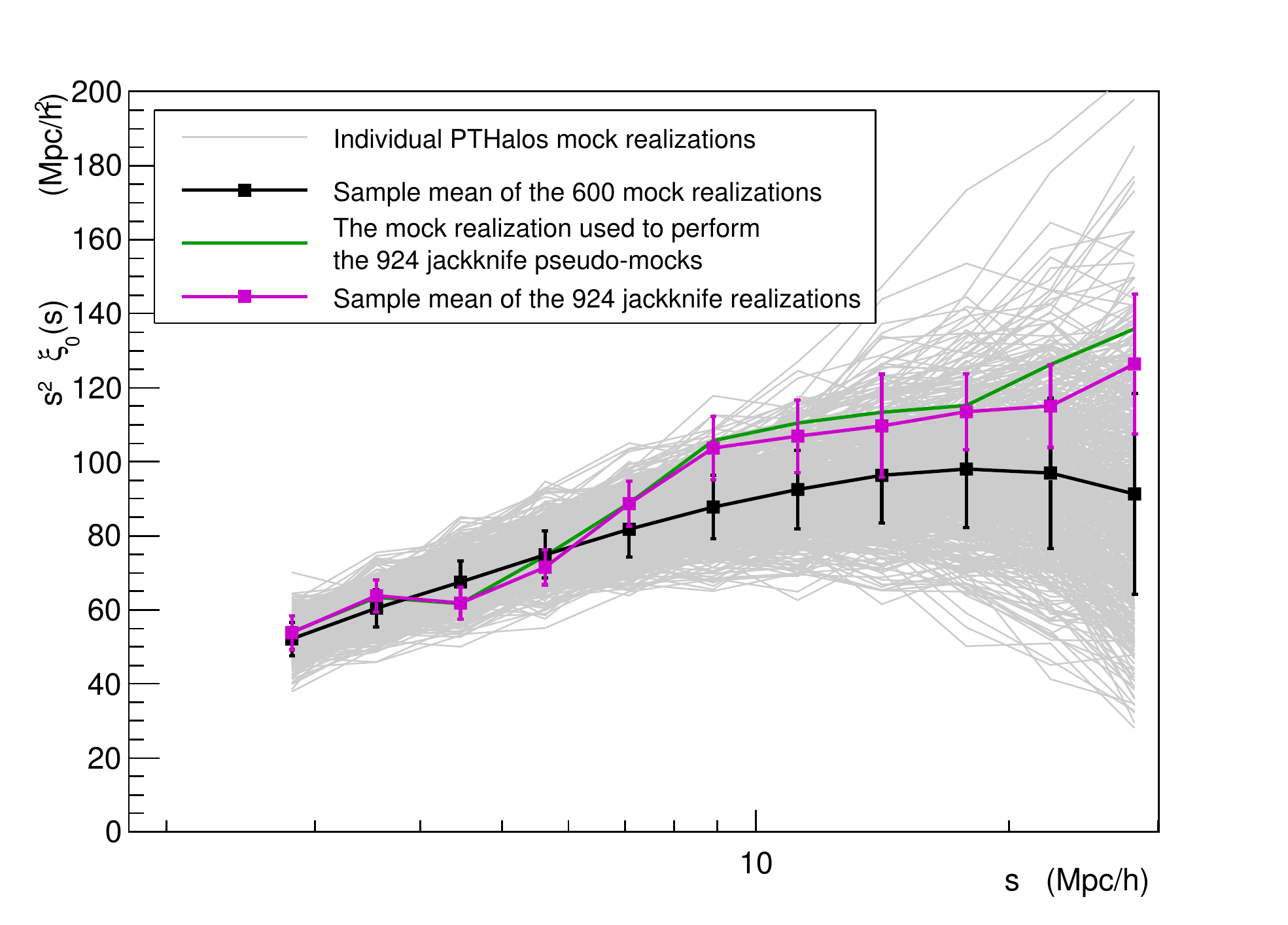}      
      \includegraphics[height=2.5in]{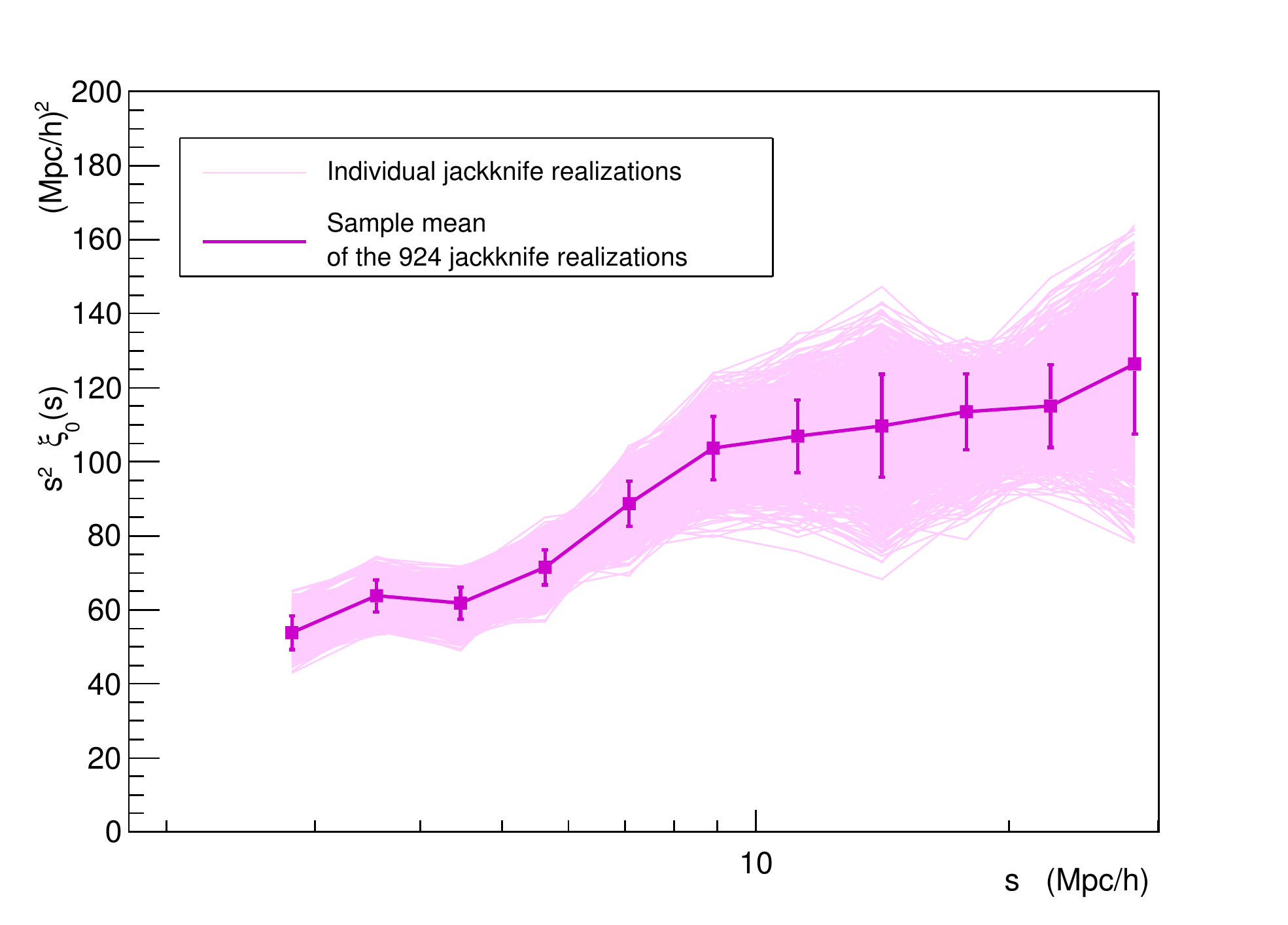}
   \end{center}   
   \caption{Monopole of the correlation function, computed for each of the $600$ mock realizations (light grey in the upper panel) and for each of the 924 jackknife pseudo-mocks (light magenta in the lower panel) generated from one given \textsc{pthalos} mock realization (green). The sample means (squares) and standard deviations (error bars) are computed from the 600 mock realizations (black) and from the 924 jackknife pseudo-mocks (dark magenta).}
   \label{fig_sample_mean}
\end{figure}

In order to illustrate the sample variance computed from an ensemble of $600$ independent realizations, we display in the upper panel of Fig.~\ref{fig_sample_mean} the monopole of the 2-pt correlation function computed for each of the 600 \textsc{pthalos} mock realizations (in light grey) and the related sample mean (in black). The dispersion of the distribution can be measured by the standard deviation (error bars on black points), simply defined as the square root of diagonal elements of the covariance matrix computed from Eq.~(\ref{eq:cov_matrix}).

\begin{figure}
   \begin{center}
      \includegraphics[height=2.5in]{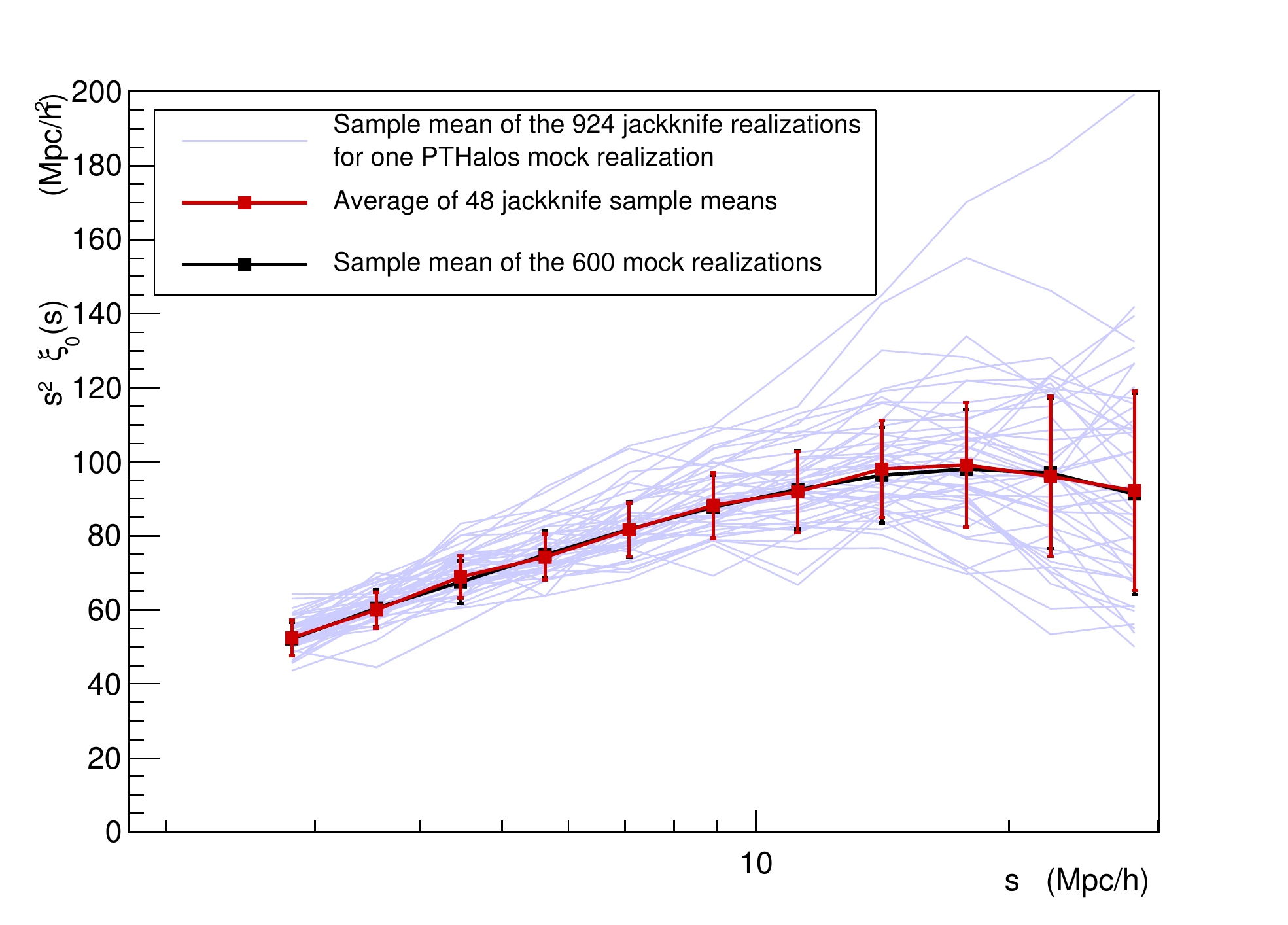}      
   \end{center}   
   \caption{Sample mean of the monopole of the correlation function for $N_M=48$ arbitrary independent mocks (in blue) for which jackknife resampling method has been applied in the ${12 \choose 6}$ configuration (924 pseudo-mocks). The average over these 48 sample means (in red) is compared to the sample mean computed directly from the 600 \textsc{pthalos} mock realizations (in black). Error bars are standard deviation of the sample mean, which are also the square root of diagonal elements of the covariance matrix.}
   \label{fig_average_mean}
\end{figure}

The lower panel of Fig.~\ref{fig_sample_mean} shows the monopole of the 2-pt correlation function computed for each of the 924 jackknife pseudo-realizations (in light magenta) generated from one single mock in the ${12 \choose 6}$ configuration (the discussion about the choice of the jackknife parameters is done in the next section).
The attractive point of the jackknife resampling technique is that only one \textsc{pthalos} mock realization is self-sufficient to define a sample variance. The sample standard deviation (error bar on dark magenta squares) is defined in this case as the square root of diagonal terms of the covariance matrix given by Eq.~(\ref{eq:cov_matrix_JK}). 

The sample mean and its standard deviation on the monopole computed from the 924 jackknife pseudo-realizations is reported in the upper panel of Fig.~\ref{fig_sample_mean} to make easier the comparison. A close attention points out a difference between sample means from the 600 \textsc{pthalos} mock realizations (in black) and from the 924 jackknife pseudo-realizations (in magenta). Actually, the latter has been computed using one single \textsc{pthalos} mock catalog (in green) and we see a very good consistency between the initial monopole function and the sample mean of the monopole function computed with the jackknife technique as we could expect.

%

We understand intuitively that if we want to faithfully reproduce the sample mean of the 600 \textsc{pthalos} mocks, we must consider many independent mock realizations. Fig.~\ref{fig_average_mean} shows the average mean over an arbitrary set of $N_M=48$ jackknife sample means, where each jackknife sample mean is the mean of $924$ jackknife pseudo-mocks, and where the standard deviation (error bars) is the square root of diagonal elements of covariance matrix as given by Eq.~(\ref{eq:cov_matrix_average_JK}). The agreement of sample means and standard deviations between the 600 \textsc{pthalos} mocks and the average over 48 mocks is very good, especially since only 48 mocks were used for this exercise. The estimate of sample means agrees within $2\%$. 

We will show in next sections that the variance is also in very good agreement, and converges faster with few independent mocks when using jackknife resampling method.

\subsection{Settings of jackknife parameters}
\label{sec:smc_param}
From this section we put aside the sample mean introduced for illustration in Section~\ref{sec:smc_mean} and we reframe the discussion on variance and covariance. In this section we will discuss about the choice of jackknife parameters in the delete-{\it d} method. For this purpose we only consider diagonal terms of covariance matrices of the correlation function and comparison of estimates of variances is done for one single mock realization ($N_M=1$).
Then we will show that jackknife variance depends on the choice of the initial mock catalog and that it is necessary to use several independent simulation realizations to average the sample variance, in a similar way of the sample mean. 

\begin{figure}
   \begin{center}
      \includegraphics[height=2.5in]{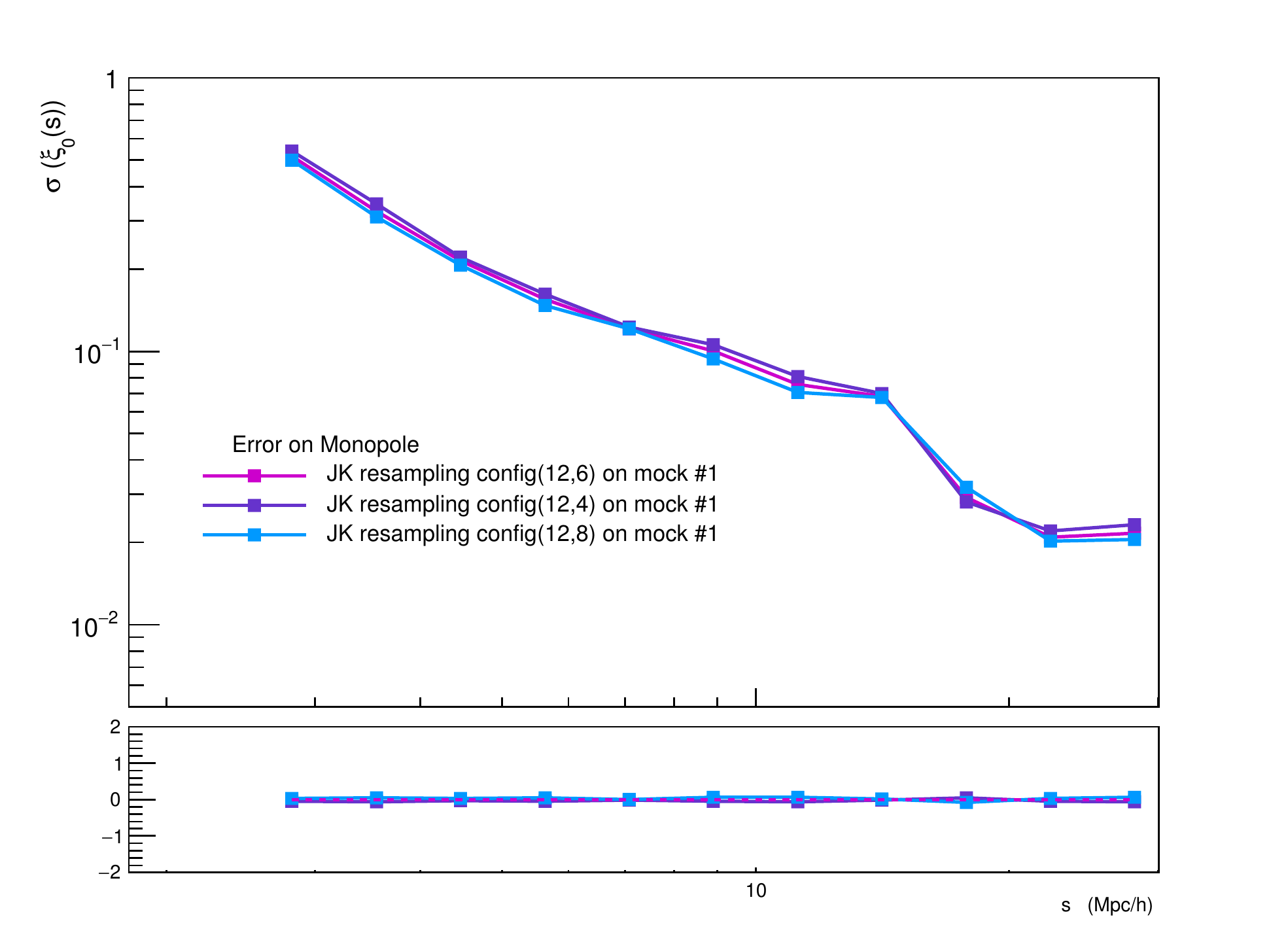}
      \includegraphics[height=2.5in]{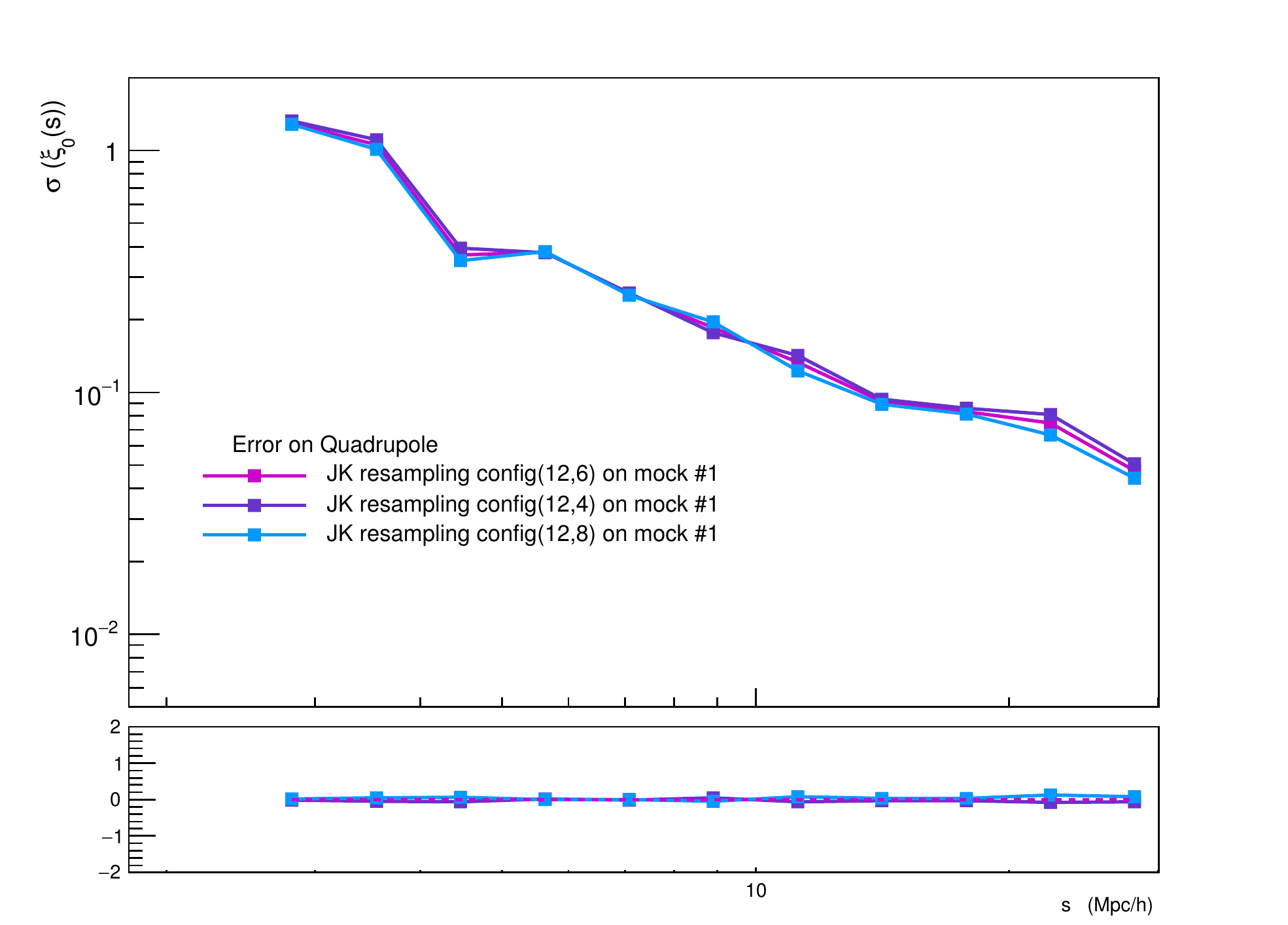}      
   \end{center}   
   \caption{Effect of the number of deleted subsamples ($N_d$) in the standard deviation estimate of the correlation function, for monopole (top panel) and quadrupole (bottom panel). Lower sub-panels display the relative difference in respect to the ${12 \choose 6}$ configuration.}
   \label{fig_nb_deleted}
\end{figure}

\begin{figure}
   \begin{center}
      \includegraphics[height=2.5in]{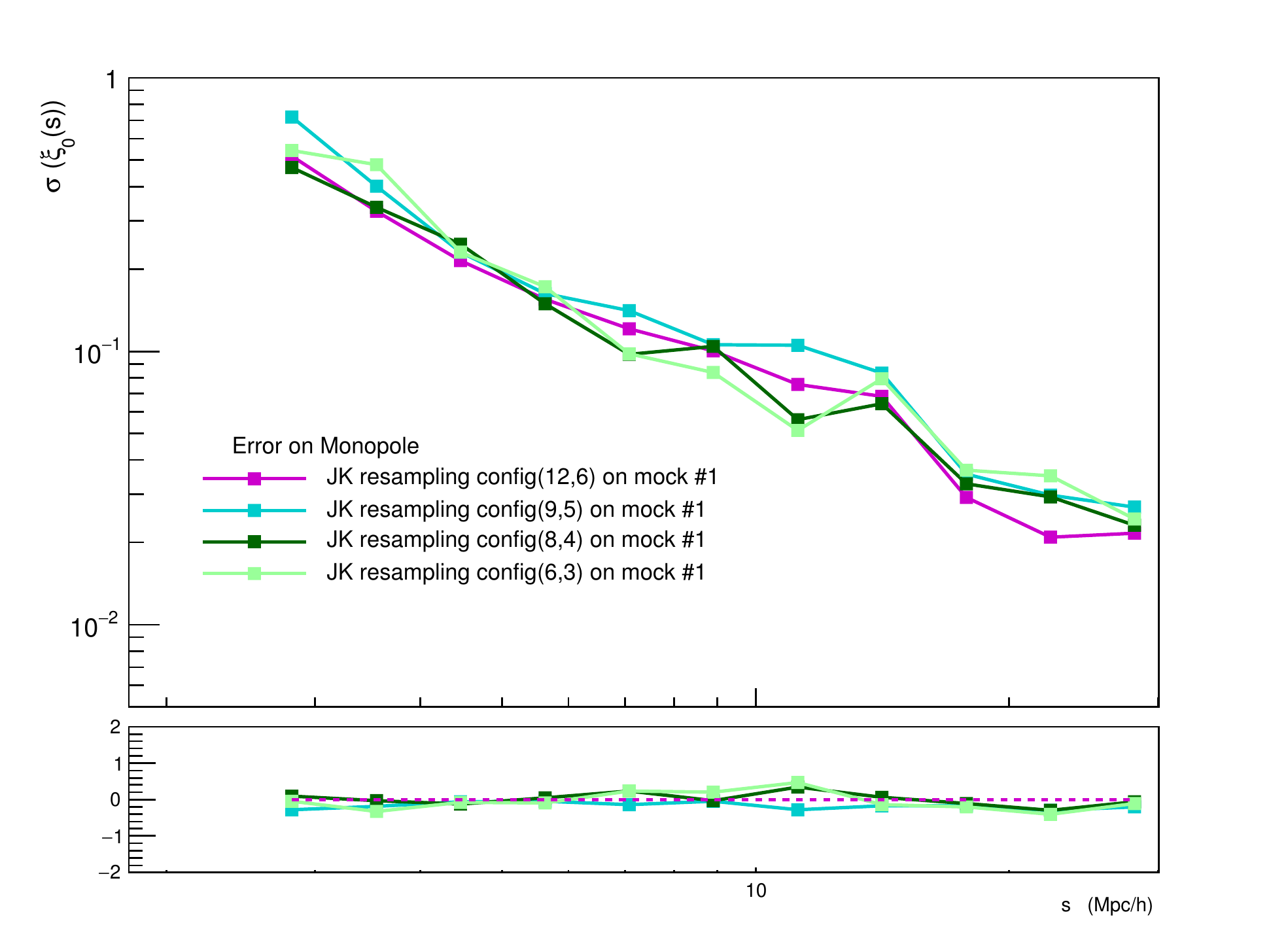}
      \includegraphics[height=2.5in]{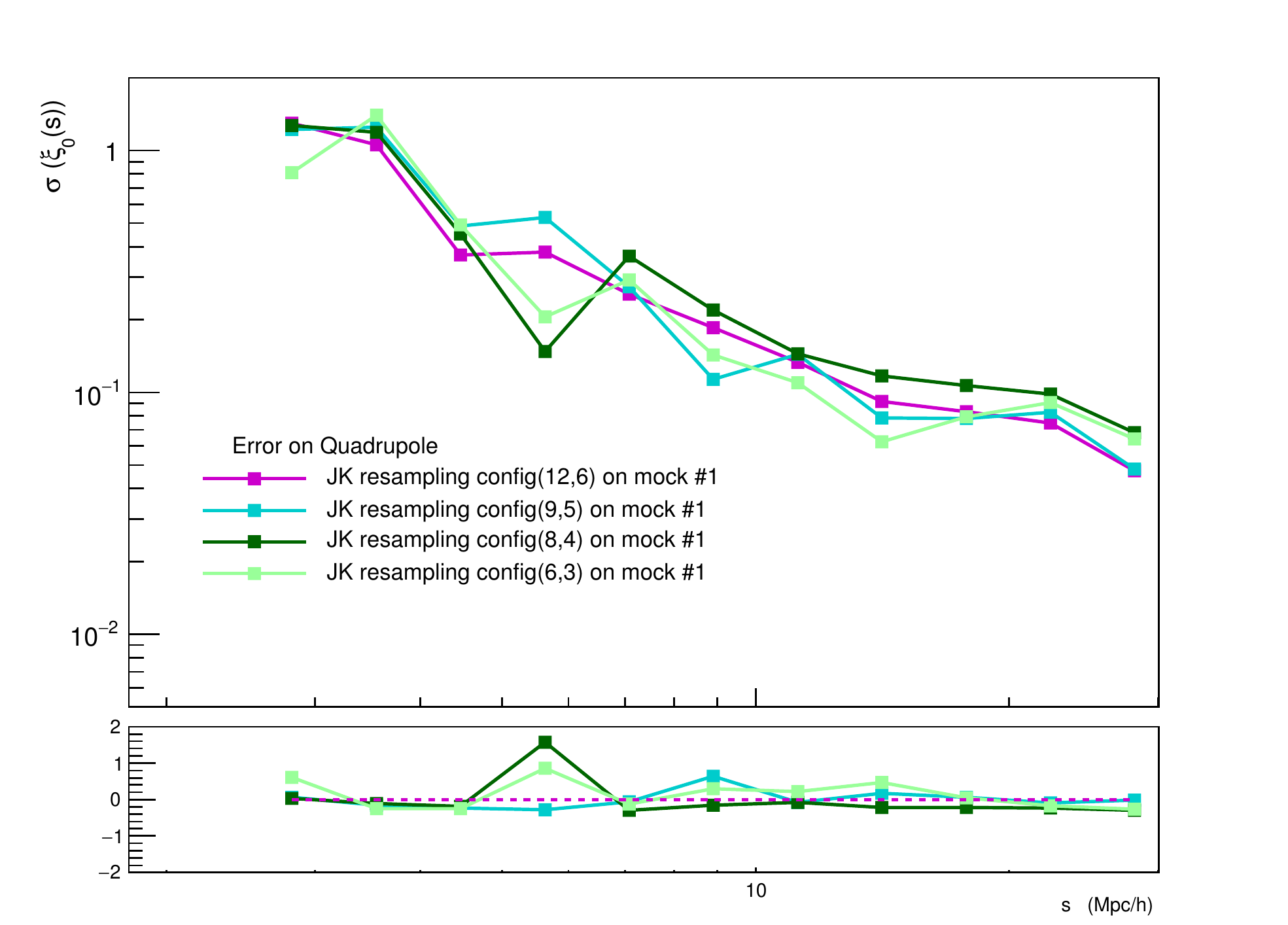}      
   \end{center}   
   \caption{Effect of the number of initial subsamples ($N_s$) in the standard deviation estimate of the correlation function, for monopole (top panel) and quadrupole (bottom panel). Lower sub-panels display the relative difference in respect to the ${12 \choose 6}$ configuration.}
   \label{fig_boxe_size}
\end{figure}

\subsubsection{Choice of ($N_s,N_d$)}
All jackknife configurations tested here are summarized in Table~\ref{tab_JKnumber}. The set of ($N_s,N_d$) values has been chosen to respect the condition given by Eq.~\ref{eq:Nd}. Given that we performed our work on the CFHTLS W1 field ($8\times9$ sq. deg.), we deliberately limited the maximum number of boxes to $12$ in order to preserve sample variance in larger scales, up to $35\;h^{-1}$Mpc.

Fig.~\ref{fig_nb_deleted} shows the influence of the number of deleted subsamples $N_d$ when the initial number of subsamples is fixed to $N_s=12$. The top and bottom panels exhibit the monopole and the quadrupole respectively, for the three jackknife configurations ${12 \choose 4}$, ${12 \choose 6}$ and ${12 \choose 8}$. The agreement between different configurations is excellent, knowing that this deviation is get from one single mock.

Fig.~\ref{fig_boxe_size} is related to the influence of the total number of subsamples $N_s$ used in the jackknife error estimate. We consider here configurations where $N_d \sim N_s/2$, for which the number of jackknife combinations is optimal for a given $N_s$. The top and bottom panels shows the monopole and quadrupole respectively, for $N_s= 6$, 8, 9 and 12 total subsamples. The modification of the number of subsamples encompasses two effects: the minimal transverse size of the subsample and the number of jackknife pseudo-realizations, which drops to only 20 realizations in the ${6 \choose 3}$ case. The spread in the standard deviation between the four curves is larger than those between the number of deleted subsamples, however we notice that the trend is the same for all curves. The larger effect of jackknife parameter settings is observed for the ${8 \choose 4}$ configuration in comparison to the ${12 \choose 6}$ case. We will show in the next section that this effect lessened after averaging over several independent mock realizations. 

\subsubsection{Initial mock realization}
\begin{figure}
   \begin{center}
      \includegraphics[height=2.5in]{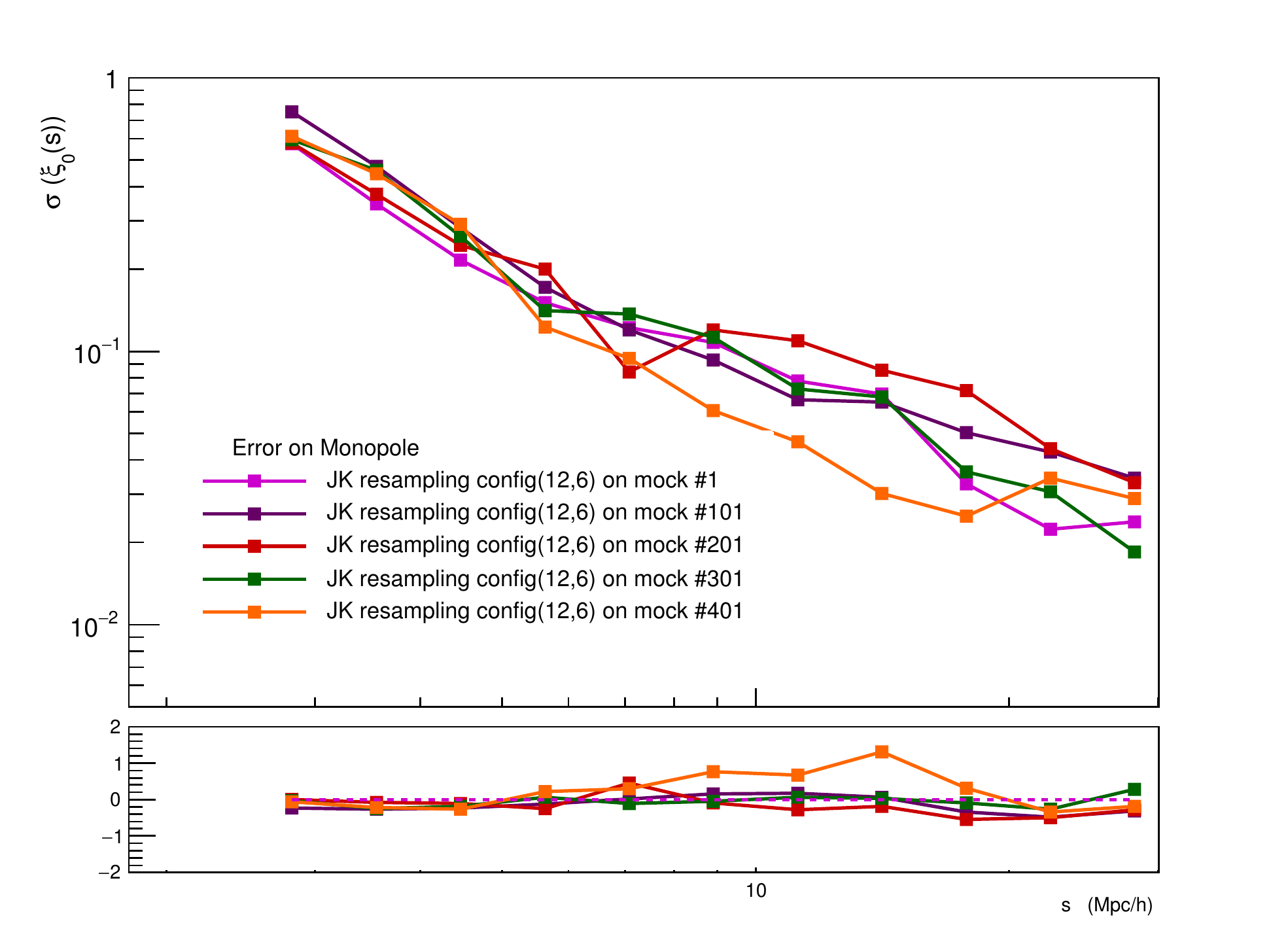}
      \includegraphics[height=2.5in]{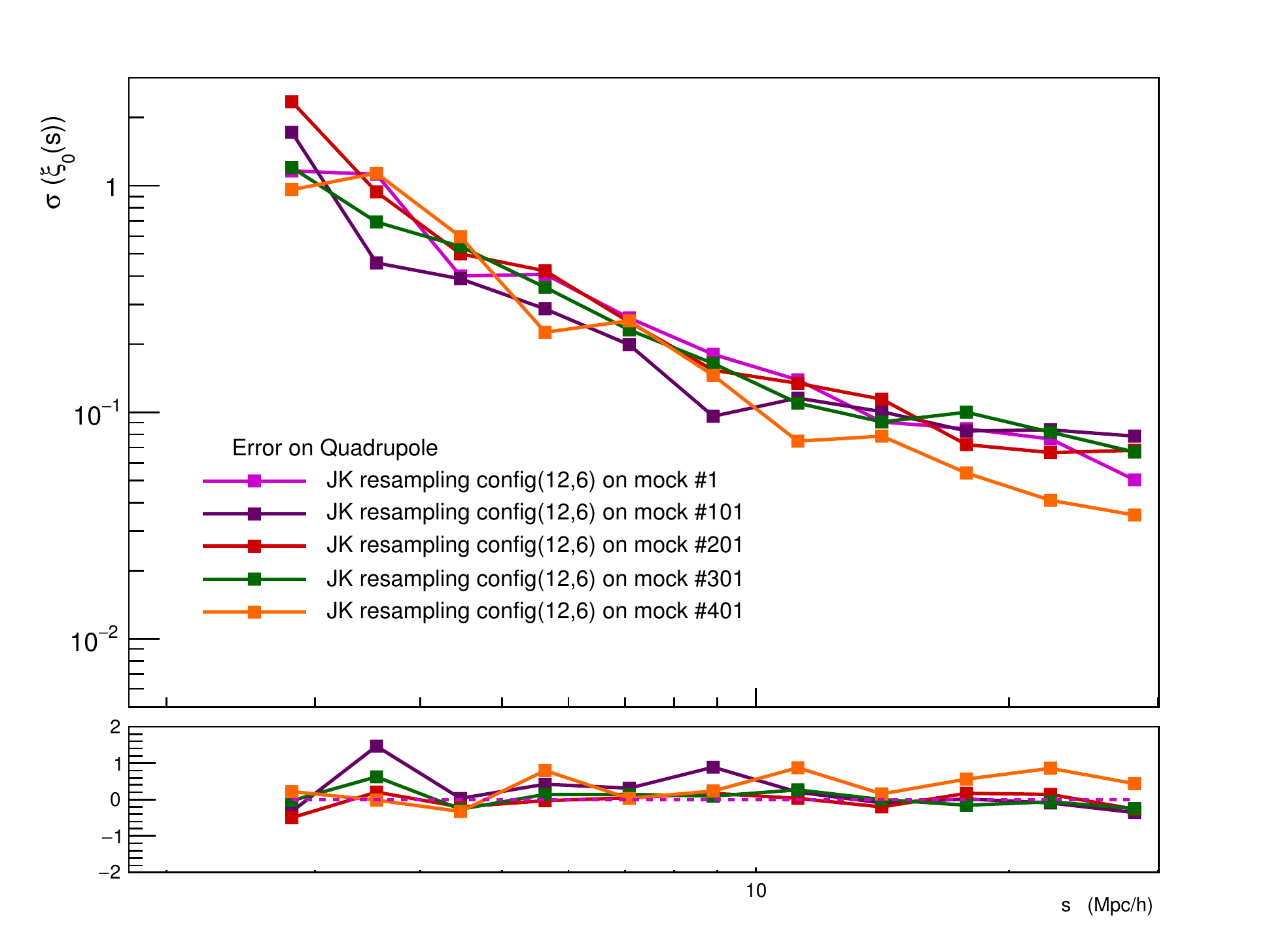}      
   \end{center}   
   \caption{Effect of the initial mock realization in the standard deviation estimate of the correlation function, for monopole (top panel) and quadrupole (bottom panel). Lower sub-panels display the relative difference in respect to the initial mock denoted $\#1$.}
   \label{fig_mock_ini}
\end{figure}

Finally we investigate the impact of the initial mock realization chosen for the jackknife resampling on the error measurement. The top and bottom panels of Fig.~\ref{fig_mock_ini} show the standard deviations computed on the monopole and quadrupole moments of the correlation function respectively, using the $12 \choose 6$ jackknife resampling method on five different initial mock realizations. Disparities between curves are large, and it is indisputably the larger effect among those studied here. However such a result was expected as each mock realization is an independent realization of the data and expresses the sample variance. 
So the initial mock realization affects not only the sample mean as seen in Section~\ref{sec:smc_mean}, but also the sample variance. One way to overcome the dependence of the initial mock is to average jackknife variances over a number of independent mocks $N_M$. 
Actually it is the strength of the jackknife method we propose here: i) the dispersion observed between each independent realization ensures that the jackknife technique gives a representative sample if applied on a sufficient number of mocks; ii) the power of the method relies on the fact that the required number of initial mocks $N_M$ is much lower than reference method with $600$ independent mocks, for a similar precision level in the sample covariance, as we will show in the next section.

\subsection{Covariance matrix}
\label{sec:smc_matrix}
Settings of jackknife parameters seem not to have strong influence on the sample variance, unlike the choice of the initial simulation realization. 
We study here the full covariance matrix estimate in case of $N_M$ independent mocks are used for the jackknife resampling and perform comparison with the covariance matrix estimate computed with $600$ independent mocks. In the following we arbitrary apply $N_M=48$. Discussion about the optimal number of $N_M$ independent realizations is addressed further down. 

\begin{figure}
   \begin{center}
      \includegraphics[height=2.5in]{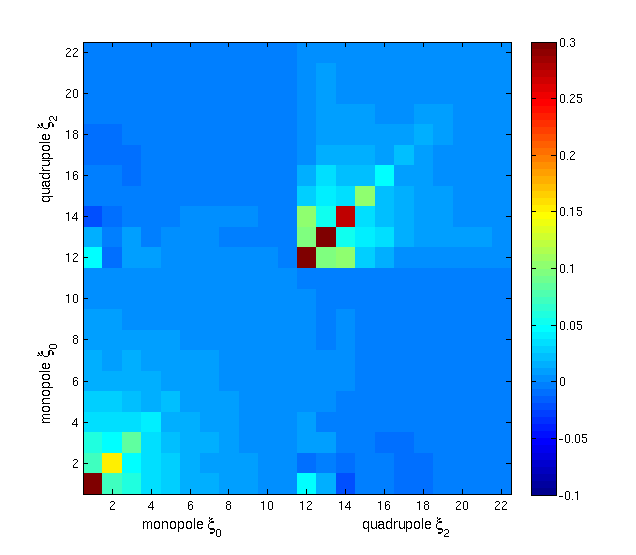}
      \includegraphics[height=2.5in]{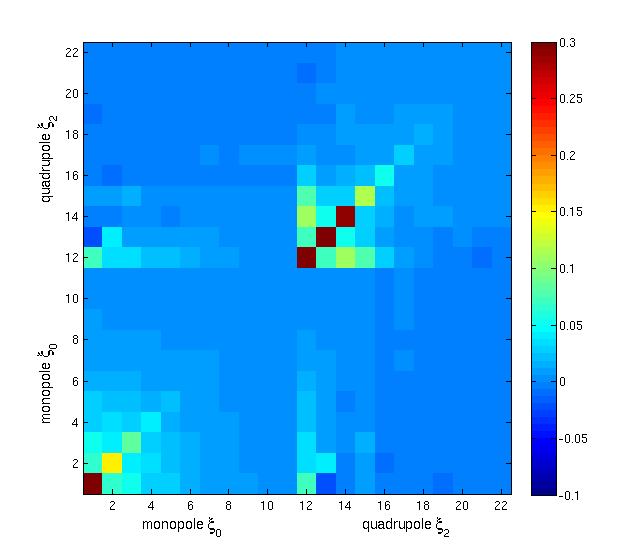}      
   \end{center}   
   \caption{Covariance matrices computed from $600$ independent mocks (top panel) and from jackknife resampling technique applied on $48$ mocks (bottom panel). Diagonal elements are displayed from the bottom left corner to the upper right. Both colorbars have same range for a better comparison.}
   \label{fig:cov_mat}
\end{figure}

The full covariance matrix $C_{ij}$ computed using $600$ independent mocks is displayed in the upper panel of Fig.~\ref{fig:cov_mat}, where bins $i=1,11$ denote the monopole of the correlation function and bins $i=12,22$ denote the quadrupole. The averaged jackknife covariance matrix $\overline{C}_{ij}$ computed from $48$ independent realizations in the ${12 \choose 6}$ jackknife configuration ($924$ pseudo-realizations for each independent mocks) is shown in the lower panel of Fig.~\ref{fig:cov_mat}. Both matrices seem to be in good agreement. The detailed comparison is performed for diagonal and off-diagonal terms separately.

\subsubsection{Diagonal elements}
\begin{figure}
   \begin{center}
      \includegraphics[height=2.5in]{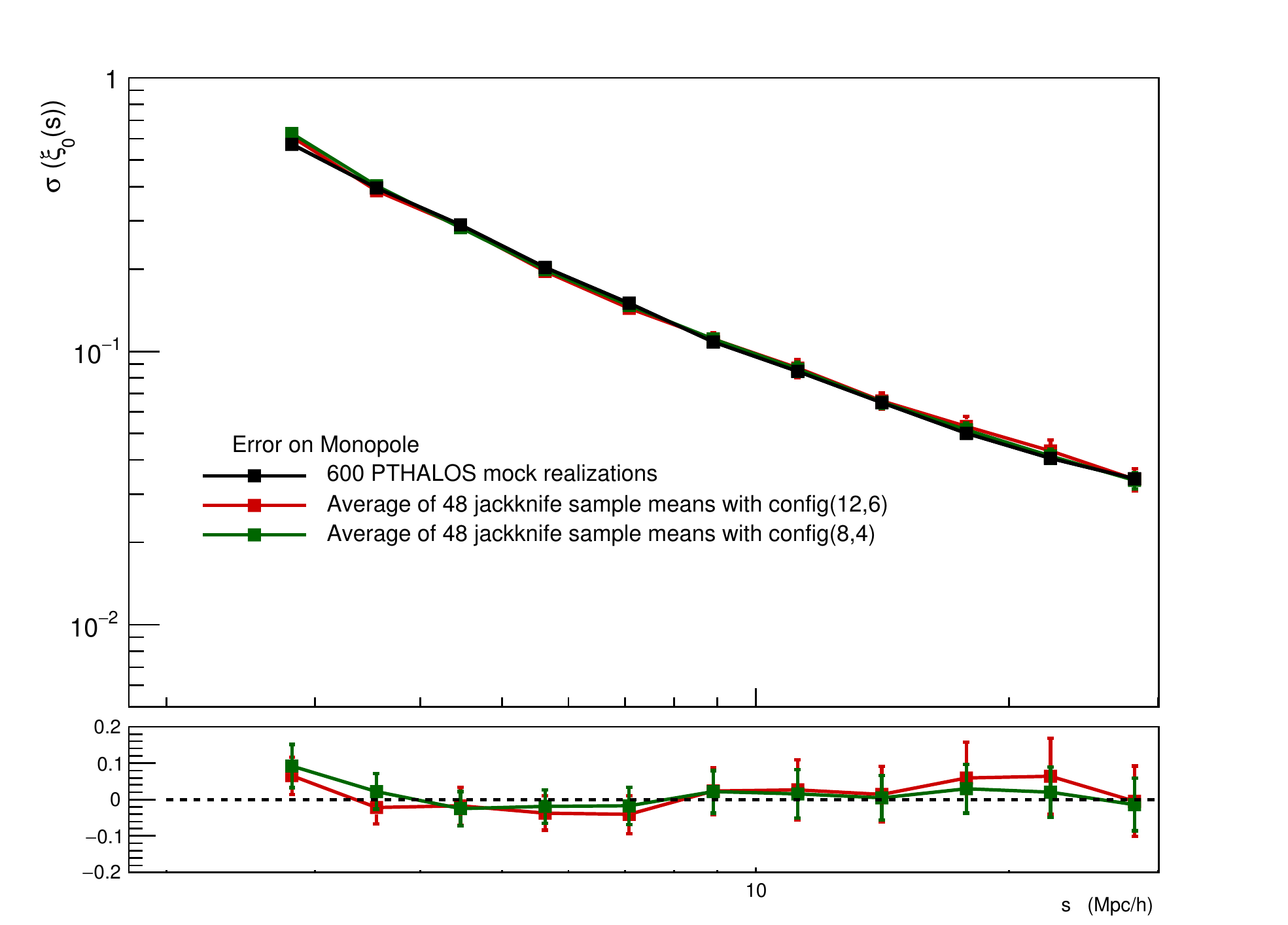}    
      \includegraphics[height=2.5in]{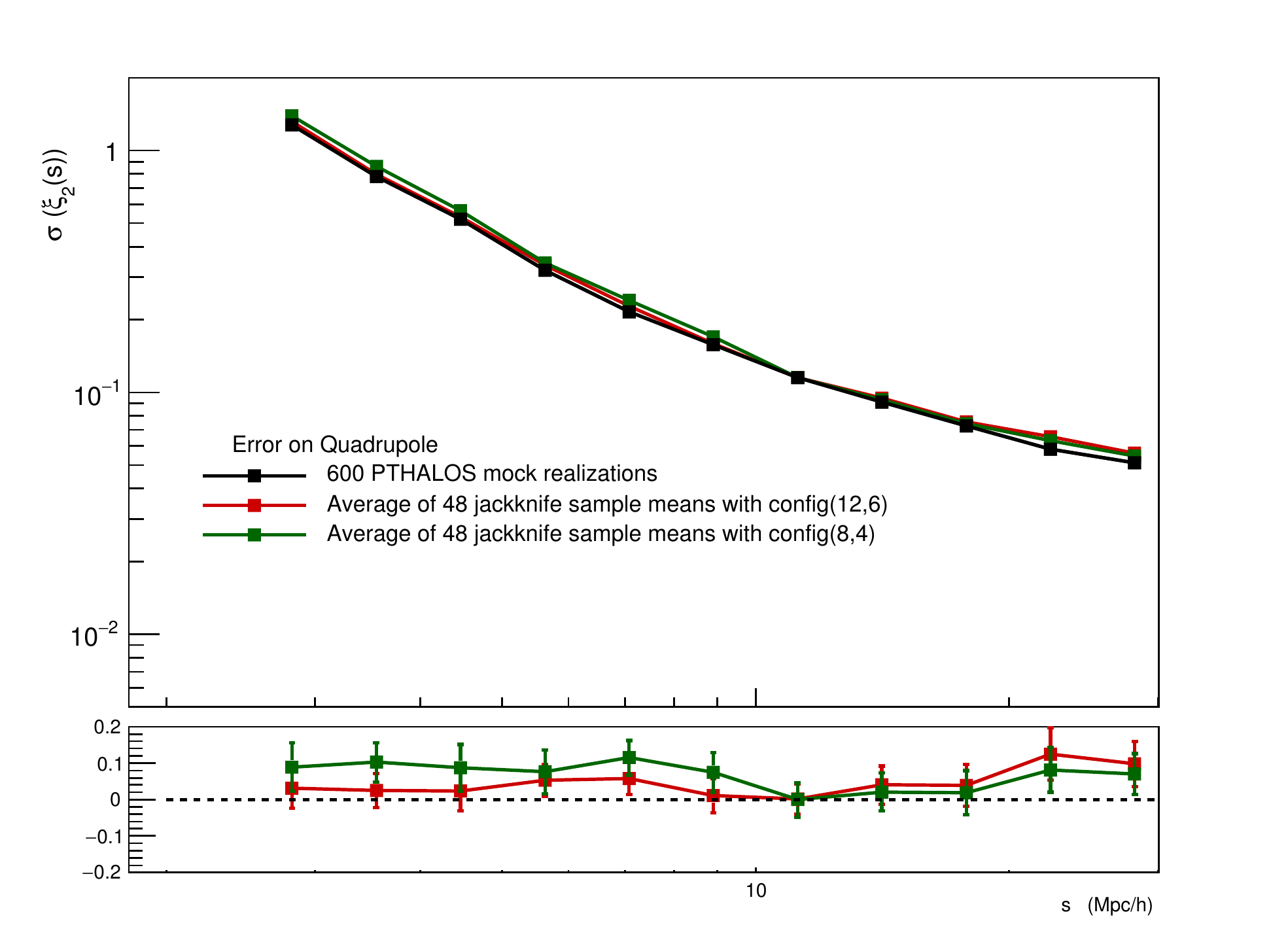}    
   \end{center}   
   \caption{Comparison of standard deviations (square root of diagonal terms of covariance matrix) on the monopole (upper panel) and quadrupole (lower panel) moments of the correlation function, between the 600 \textsc{pthalos} mocks case and the jackknife resampling case applied on 48 mocks. Lower sub-panels display the relative difference in respect to the 600 \textsc{pthalos} mocks case.}
   \label{fig:var}
\end{figure}

We report in Fig.~\ref{fig:var} the standard deviation of the monopole and quadrupole of the correlation function, defined as the square root of diagonal elements of the covariance matrix ($\sqrt{{C}_{ii}}$), for the reference case of 600 \textsc{pthalos} mocks and for the special case of jackknife resampling over 48 initial mocks with the ${12 \choose 6}$ configuration. 
The error bars in Fig.~\ref{fig:var} are error on standard deviation. In the case of \textsc{pthalos} mocks this error matches with the standard deviation on standard deviation:
\begin{equation}
  \sigma[\sqrt{{C}_{ii}}] =  \frac {\sigma[{C}_{ii}]} {2  \sqrt{{C}_{ii}}} =  \frac {\sqrt{{C}_{ii}}} {\sqrt{2(N_s-1)}}
\end{equation}
with the variance of each elements of the covariance matrix $\sigma[{C}_{ii}]$ given by Eq.~\ref{eq:cov_matrix_var}. 

In the special case of jackknife resampling over 48 initial mocks, we define the error on the standard deviation as:
\begin{equation}
  \sigma[\sqrt{\overline{C}_{ii}}] =  \frac {\sigma[\sqrt{\widehat{C}_{ii}}]} {\sqrt{N_M}} =  \frac {\sigma[\widehat{C}_{ii}]} {2  \sqrt{N_M \overline{C}_{ii}}} 
\end{equation}
where $\sigma[\widehat{C}_{ii}]$ is defined according to Eq.~\ref{eq:variance_average_JK}.

For the jackknife resampling, we report also in Fig.~\ref{fig:var} the average standard deviations for the ${8 \choose 4}$ configuration, since it was the larger effect of jackknife parameter settings observed in Section~\ref{sec:smc_param}. Finally, the relative difference between 600 \textsc{pthalos} mocks and jackknife resampling over 48 initial mocks is displayed in lower sub-panels, showing that the agreement is good. 

To further investigate the concordance, we apply a Fisher-Snedecor test~\citep{Snedecor:1989} which is known as the F-test to test the equality of the variances of two populations. Under the null hypothesis :
\begin{equation}
  H_0:   {\sigma_1^2} = {\sigma_2^2} 
\end{equation}
the F-test consists in calculating the ratio:
\begin{equation}
  F =  \frac {s_1^2} {s_2^2} \frac {\sigma_2^2} {\sigma_1^2} = \frac {s_1^2} {s_2^2} 
\end{equation}
where $s^2_i$ are the sample variances (${C}_{ii}$) and $\sigma^2_i$ are the true variances of the sample $i$.

The null hypothesis is rejected if:
\begin{equation}
F<F_{1-\alpha/2} (N_1-1, N_2-1)
\end{equation}
or 
\begin{equation}
F>F_{\alpha/2} (N_1-1, N_2-1) 
\end{equation}
where $F_{\alpha/2} (N_1-1, N_2-1)$ is the critical value of the F distribution with $N_1-1$ and $N_2-1$ degrees of freedom at a significance level $\alpha$. 

As the F-test should be applied on independent populations, we compare variances from the jackknife resampling method on 48 mocks with variances from the reference case using the remaining 552 \textsc{pthalos} mocks over the $600$ initial mocks. In this case, the rejection region at the significance level $\alpha=0.05$ is:
\begin{equation}
F< F_{0.975} (551, 47) = 0.685 
\end{equation}
or
\begin{equation}
 F>F_{0.025} (551, 47) = 1.57
\end{equation}

Table~\ref{tab_Ftest} summarizes results of the F-test and shows that there is not enough evidence to reject the null hypothesis at the $0.05$ significance level for any diagonal terms in the monopole and quadrupole moments.


\subsubsection{Off-diagonal elements}
In order to study off-diagonal terms of the covariance matrix, we define the reduced covariance matrix or the correlation matrix such as:
\begin{equation}
  r_{ij} = \frac {C_{ij}} {\sqrt{C_{ii}C_{jj}}}
\end{equation}

\begin{figure}
   \begin{center}
      \includegraphics[height=2.5in]{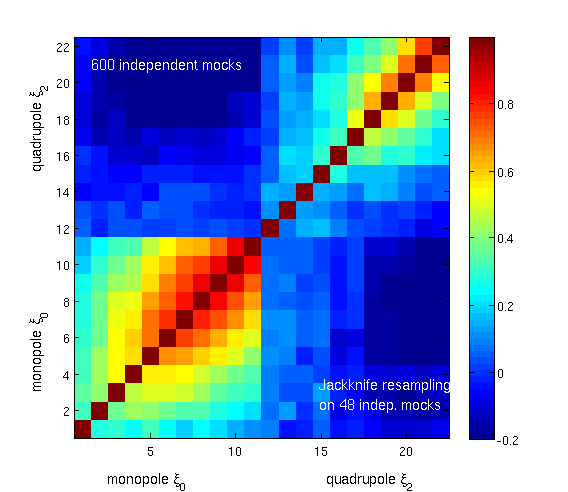}    
   \end{center}   
   \caption{Comparison of the correlation matrix $\frac {C_{ij}}{\sqrt{C_{ii}C_{jj}}}$ between the standard estimation using $600$ independent realizations  (upper triangular part) and the estimation used in this work using jackknife resampling technique on $48$ initial independent realizations (lower triangular part). Diagonal elements from the bottom left corner to the upper right are common to both estimates and are everywhere equal to 1.}
   \label{fig:corr_mat}
\end{figure}

The upper triangular of Fig.~\ref{fig:corr_mat} shows the correlation matrix from the standard estimate using the $600$ \textsc{pthalos} independent mocks, while the lower triangular part shows the correlation matrix from the jackknife estimate. The diagonal from the bottom left corner to the upper right is common to both estimates and is everywhere equal to 1.
Comparison between upper part and lower part of Fig.~\ref{fig:corr_mat} shows that the agreement between correlation coefficients is quite very good.

In conclusion, the full covariance matrix estimate computed with $600$ mock realizations or with the jackknifed average over arbitrary $48$ mock realizations
seems to give similar values.

\subsection{Precision matrix}
\label{sec:smc_precision}
We test in a similar way the comparison of precision matrices. In the case of the covariance matrix has been estimated from an ensemble of $N_s$ simulation mocks, we get a raw precision matrix, computed as $\widehat{C}_{ij}^{-1}$, and an unbiased $\Psi_{ij}$, corrected from the Whishart bias according to Eq.~\ref{eq:hartlap_mean}. In the jackknife case, the precision matrix is simply defined as the inverse of the averaged covariance matrix as defined by Eq.~\ref{eq:prec_matrix_JK}. Like for covariance matrix, we examine diagonal terms and off-diagonal terms separately.  

\begin{figure}
   \begin{center}
      \includegraphics[height=2.5in]{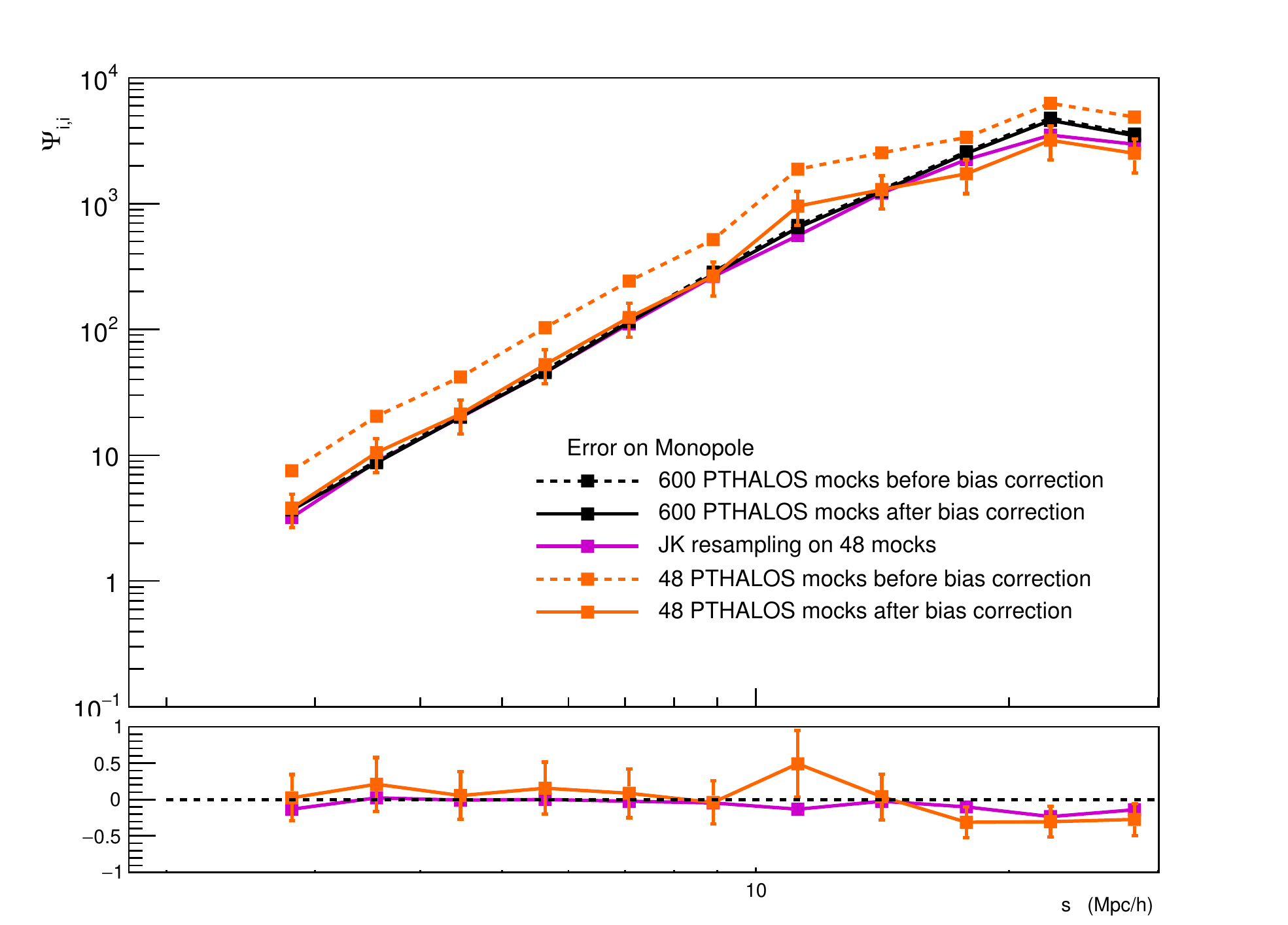}
      \includegraphics[height=2.5in]{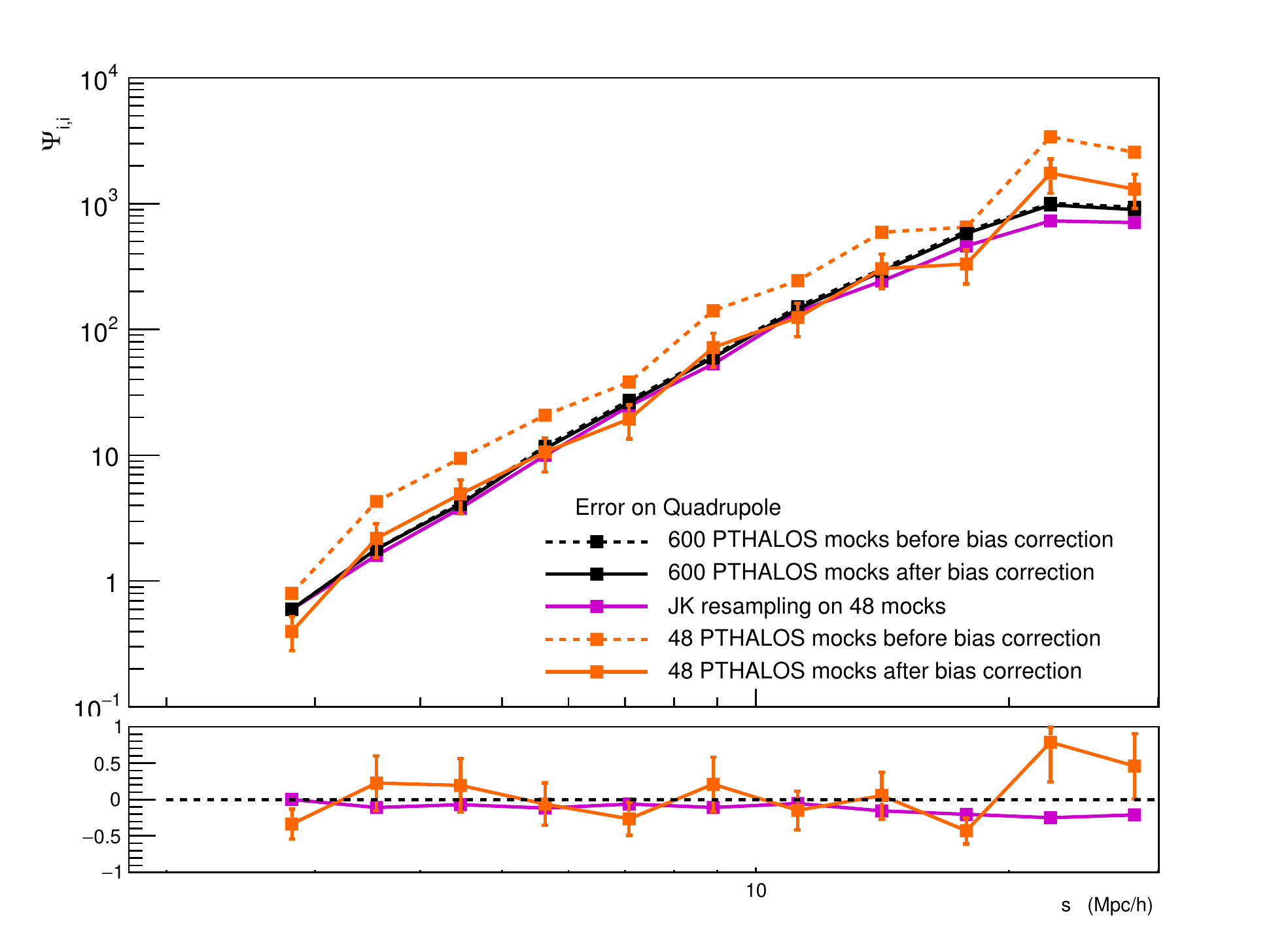}      
   \end{center}   
   \caption{Comparison of diagonal elements of the precision matrix estimate, for the monopole (upper panel) and the quadrupole (lower panel), between 600 \textsc{pthalos} mocks, the jackknife resampling applied on 48 mocks and 48 mocks without resampling. Lower sub-panels display the relative difference in respect to the 600 \textsc{pthalos} mocks case corrected from the Whishart bias.}
   \label{fig:prec_diag}
\end{figure}

\subsubsection{Diagonal elements}
Fig.~\ref{fig:prec_diag} shows the diagonal elements of the precision matrix for the monopole and quadrupole moments of the correlation function. For the reference case of 600 \textsc{pthalos} mocks, we report the estimate of the precision matrix before and after the bias correction introduced in Section~\ref{sec:cov_mock}. In this figure is also reported the estimate from the jackknife resampling over $48$ mocks, as well as the estimate from $48$ mocks without any resampling. The error bars are only displayed for precision matrix terms after the Whishart bias correction, where $\sigma(\Psi)$ is defined according to Eq.~\ref{eq:hartlap_var}.
We note that the precision matrix estimate with the jackknife resampling seems to be in good agreement with those from $600$ \textsc{pthalos} mocks. The agreement of diagonal terms between $600$ and $48$ mocks seems also reasonable, but only after the Whishart correction and with a larger dispersion.

\begin{figure}
   \begin{center}
      \includegraphics[height=2.5in]{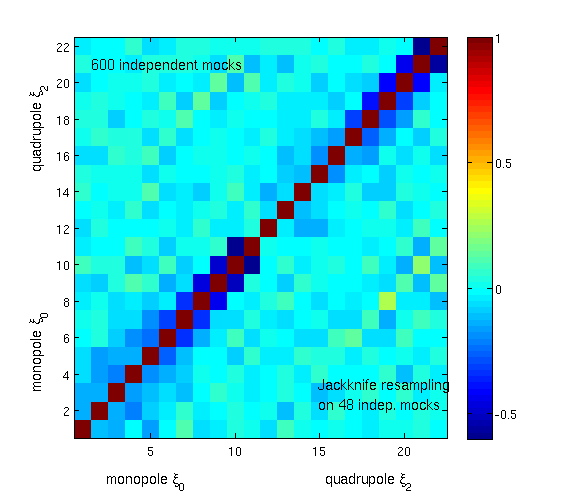}
      \includegraphics[height=2.5in]{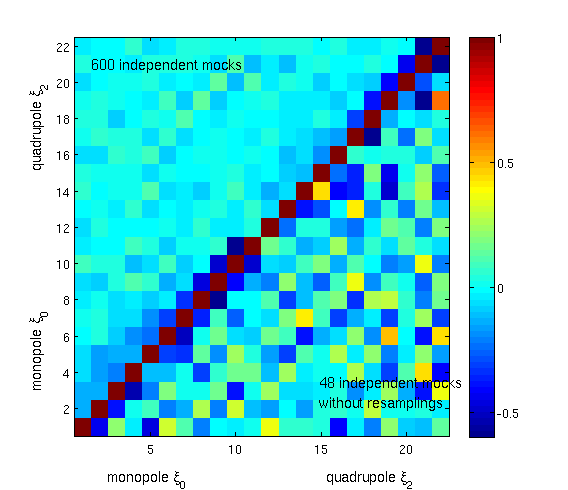}      
   \end{center}   
   \caption{Top panel: Reduced precision matrix $\frac {\Psi_{ij}}{\sqrt{\Psi_{ii}\Psi_{jj}}}$  computed from the standard estimate using $600$ independent mocks (upper triangular part) and from the jackknife resampling technique applied on $48$ initial mocks (lower triangular part). Bottom panel: Same reduced precision matrix for which the lower triangular part is the estimate using $48$ independent mocks without any resamplings. Diagonal elements from the bottom left corner to the upper right are common to both estimates and are everywhere equal to 1.}
   \label{fig:prec_mat}
\end{figure}

\subsubsection{Off-diagonal elements}
In order to compare off-diagonal terms, we compute the reduced precision matrix $\frac {\Psi_{ij}}{\sqrt{\Psi_{ii}\Psi_{jj}}}$ for which all diagonal elements are equal to unity. 
The top panel of Fig.~\ref{fig:prec_mat} shows the comparison between $600$ independent realizations (upper triangular part) and the jackknife resampling technique on $48$ independent mocks in the ${12 \choose 6}$ configuration (lower triangular part). The bottom panel of Fig.~\ref{fig:prec_mat} gives the comparison between $600$ independent realizations (upper triangular part) and $48$ independent mocks without any resampling (lower triangular part). 
This example highlights the convergence power of the jackknife method, mainly on the correlation coefficients.

\subsection{Convergence rate}
\label{sec:convergence}
The \textsc{smc} method seems to offer a good alternative to estimate sample variances from an ensemble of few mock realizations.  All the discussion along this paper deals with an arbitrary choice of $N_M=48$ mocks used for the jackknife resample. In this section we are interested in the accuracy of the variance and we estimate the equivalent number of mock catalogues $N_M$ needed to yield to a similar accuracy on variance than if computed with $N_s$ simulation mocks.

Taking advantage of properties of the trace of a matrix, we compare the relative uncertainty of the standard deviation of the covariance matrix normalized over all diagonal elements. In the case of simulation covariance, this relative uncertainty can be written as:
\begin{equation}
  \frac {\sum\sigma[\sqrt{{C}_{ii}}]} {Tr(\sqrt{{C}_{ii}})} =  \frac {1} {\sqrt{2(N_s-1)}}
   \label{eq:trace_MM}
\end{equation}

While in the jackknife case the relative accuracy of the standard deviation is:
\begin{equation}
  \frac {\sum\sigma[\sqrt{\overline{C}_{ii}}]} {Tr(\sqrt{\overline{C}_{ii}})} =  \frac {1} {\sqrt{2N_M}} \frac {\sum\sigma[\overline{C}_{ii}]} {Tr(\overline{C}_{ii})} 
   \label{eq:trace_JK}
\end{equation}

\begin{figure}
   \begin{center}
      \includegraphics[height=2.5in]{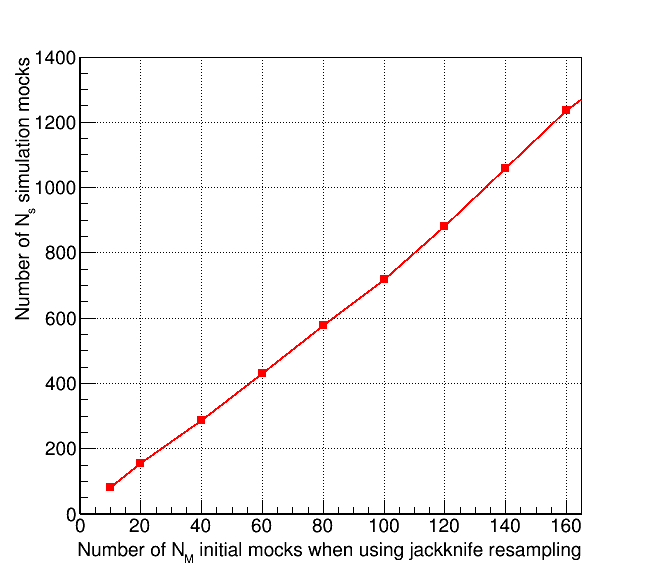}    
   \end{center}   
   \caption{Reduction in the number of simulations when using the jackknife resampling method over $N_M$ initial mocks needed to give a similar accuracy in the covariance matrix estimate than those computed from $N_s$ simulation mocks.}
   \label{fig:calcul_ns}
\end{figure}

The identification of Eq.~\ref{eq:trace_MM} and Eq.~\ref{eq:trace_JK} gives the relationship between the number of initial mock realizations $N_M$ needed to give a similar accuracy of the covariance matrix estimate than those computed from $N_s$ simulation mocks, as illustrated in Fig.~\ref{fig:calcul_ns}. By comparing the rates of convergence, we show that only $N_M=85$ mocks are needed instead of $N_s=600$, meaning that the reduction in the number of mock simulations is about a factor $7$. This result is at the same order of magnitude than those obtained with the Fourier mode-resampling method~\citep{Schneider:2011}.

\section{Conclusions}
\label{sec:conclusion}

We propose in this paper a novel approach to compute sample covariance matrices with fewer mocks than usually required by Wishart statistics. We have shown that jackknife resamplings can be applied on mock catalogues and that the internal dispersion observed between each independent realization ensures that the jackknife technique gives a representative sample when applied on a sufficient number of mocks. We find the fast convergence of the method and that 
the required number of initial mocks is $7$ much lower than if computing covariance matrix estimate from an set of mock catalogues without resampling. Finally we expect that this alternative method could be applied for cosmological analysis in case of a few N-body simulations available. 

\section*{Acknowledgements}
We thank Stephane Placzynski for useful discussion. 
The mocks used were produced in SCIAMA High Performance Supercomputer (HPC) cluster, supported by the ICG, SEPNet and the University of Portsmouth.
SE and AP acknowledge financial support from the grant OMEGA ANR-11-JS56-003-01.
AP and SLT acknowledge the support of the OCEVU Labex (ANR-11-LABX-0060) and the A*MIDEX  project (ANR-11-IDEX-0001-02) funded by the 'Investissements  d'Avenir' French government program managed by the ANR.




\bibliographystyle{mnras}
\bibliography{jackknife}




%
%
\onecolumn
\begin{table}
    \caption{Number of jackknife realizations $N_{\textsc{jk}}$ for all possible configurations leaving out $N_d$ sub-samples amongst $N_s$ initial sub-volumes. The size of individual box is shown on an indicative basis, knowing the initial CFHT-LS Wide 1 field.}
    \begin{tabular}{c c c c}
        \hline \hline
        Number of sub-volumes  	& Number of deleted 	& Number of jackknife                    & Minimal transverse size    \\
         $N_s$ 			        &  sub-volumes $N_d$       & combinations  $N_{\textsc{jk}}$ &  at $z=0.57$ ($Mpc.h^{-1}$)\\
        \hline
        6   & 3 & 20   &48  \\
        8   & 4 & 70   & 36 \\
        9   & 5 & 126 & 35 \\
        12 & 4 & 495 & 35 \\
        12 & 6 & 924 & 35 \\
        12 & 8 & 495 & 35 \\
        \hline
    \end{tabular}
    \label{tab_JKnumber}
\end{table}

\begin{table}
    \caption{F-test for equality of two variances between the reference case with $N_1=552$ independent \textsc{pthalos} mocks and the jackknife resampling case on $N_2=48$ mocks, for the monopole and quadrupole moments of the correlation function.}
     \begin{tabular*}{\textwidth}{@{\extracolsep{\fill}} l c c c c c c c c c c c}
        \hline \hline
        bin  	&  1 & 2 & 3 & 4 & 5 & 6 & 7 & 8 & 9 & 10 & 11 \\
        \hline  \\
        \multicolumn{5}{l}{{\bf Variances ($s^2_{i}$) on monopole moment}} \\
        \textsc{pthalos} mocks   & 0.3274 & 0.1561 & 0.08418 & 0.04111 & 0.02236 & 0.01174 & 0.00715 & 0.00421 & 0.00252 & 0.00164 & 0.00117  \\
        Jackknife                       & 0.3808 & 0.1617 & 0.07487 & 0.03588 & 0.01983 & 0.01259 & 0.00686 & 0.00442 & 0.00289 & 0.00195 & 0.00132\\
        F-test                             & 0.860   & 0.965   &  1.124    &  1.146    &  1.128    &  0.9325  & 1.042   &  0.952      &  0.872    & 0.841    & 0.886\\
        \hline\\
        \multicolumn{5}{l}{{\bf Variances ($s^2_{i}$) on quadrupole moment}} \\
        \textsc{pthalos} mocks  & 1.6309 & 0.6066 & 0.2704 & 0.1021 & 0.04650 & 0.02485 & 0.01326 & 0.00832 & 0.00530 & 0.00339 & 0.00260  \\
        Jackknife                      &  1.7734 & 0.6959 & 0.2859 & 0.1216 & 0.05112 & 0.02760 & 0.01339 & 0.00945 & 0.00590 & 0.00452 & 0.00320 \\
        F-test                            &  0.920  &  0.872  &  0.946   & 0.840   &  0.909   & 0.900    & 0.990    &   0.880      &  0.898    &  0.750   &  0.813\\
        \hline \\
    \end{tabular*}
    \label{tab_Ftest}
\end{table}


\bsp	
\label{lastpage}
\end{document}